\documentclass[conference]{IEEEtran}
\IEEEoverridecommandlockouts
\usepackage{cite}
\usepackage{amsmath,amssymb,amsfonts,bm}
\usepackage{makecell}
\usepackage{algorithm}
\usepackage{algorithmic}
\usepackage{graphicx}
\usepackage{textcomp}
\usepackage{soul}
\usepackage{xcolor}
\def\BibTeX{{\rm B\kern-.05em{\sc i\kern-.025em b}\kern-.08em
    T\kern-.1667em\lower.7ex\hbox{E}\kern-.125emX}}
\begin{document}

\title{CAM/CAD Point Cloud Part Segmentation\\ via Few-Shot Learning
\thanks{This work has been submitted to the IEEE for possible publication.
Copyright may be transferred without notice, after which this version may no longer be accessible.}
}

\author{
\IEEEauthorblockN{Jiahui Wang}
\IEEEauthorblockA{\textit{Electrical and Computer Engineering}\\
\textit{National University of Singapore}\\
Singapore, Singapore\\
wjiahui@u.nus.edu}
\\
\IEEEauthorblockN{Abdullah Al Mamun}
\IEEEauthorblockA{\textit{Electrical and Computer Engineering}\\
\textit{National University of Singapore}\\
Singapore, Singapore\\
a.almamun@nus.edu.sg}
\and
\IEEEauthorblockN{Haiyue Zhu}
\IEEEauthorblockA{\textit{Singapore Institute of Manufacturing}\\
\textit{Technology (SIMTech), A*STAR}\\
Singapore, Singapore\\
zhu\_haiyue@simtech.a-star.edu.sg}
\\
\IEEEauthorblockN{Vadakkepat, Prahlad}
\IEEEauthorblockA{\textit{Electrical and Computer Engineering}\\
\textit{National University of Singapore}\\
Singapore, Singapore\\
prahlad@nus.edu.sg}
\and
\IEEEauthorblockN{Haoren Guo}
\IEEEauthorblockA{\textit{Electrical and Computer Engineering}\\
\textit{National University of Singapore}\\
Singapore, Singapore\\
haorenguo\_06@u.nus.edu}
\\
\IEEEauthorblockN{Tong Heng Lee}
\IEEEauthorblockA{\textit{Electrical and Computer Engineering}\\
\textit{National University of Singapore}\\
Singapore, Singapore\\
eleleeth@nus.edu.sg}
}
\maketitle

\begin{abstract}
3D part segmentation is an essential step in advanced CAM/CAD workflow. 
Precise 3D segmentation contributes to lower defective rate of work-pieces produced by the manufacturing equipment 
(such as computer controlled CNCs), thereby improving work efficiency and attaining the attendant economic benefits. 
A large class of
existing works on 3D model segmentation are mostly based on fully-supervised learning, 
which trains the AI models with large, annotated datasets. 
However, the disadvantage is that the resulting models from the fully-supervised learning methodology
are highly reliant
on the completeness (or otherwise) of the available dataset, 
and its generalization ability is relatively poor to new unknown/unseen segmentation types
({\it i.e.}, further additional so-called novel classes). 
In this work, we propose and develop a noteworthy few-shot learning-based approach 
for effective part segmentation in CAM/CAD; 
and this is designed 
to significantly enhance its generalization ability,
and our development also
aims to flexibly adapt to new segmentation tasks by using only relatively rather few samples. 
As a result, it not only reduces the requirements for the
usually unattainable and
exhaustive completeness of supervision datasets, 
but also improves the flexibility for real-world applications. 
In the development, drawing inspiration from the pertinent and interesting work described in the open literature as the attMPTI network, 
we propose and develop a multi-prototype approach (with self-attention mechanics) for few-shot point cloud part segmentation. 
As further improvement and innovation, we additionally adopt the transform net and the center loss block in the network. 
These characteristics serve to
improve the comprehension 
for 3D features of the various possible instances of the whole work-piece
and ensure the close distribution of the same class in feature space. 
\end{abstract}


\begin{figure}
\centering
\includegraphics[scale=0.63]{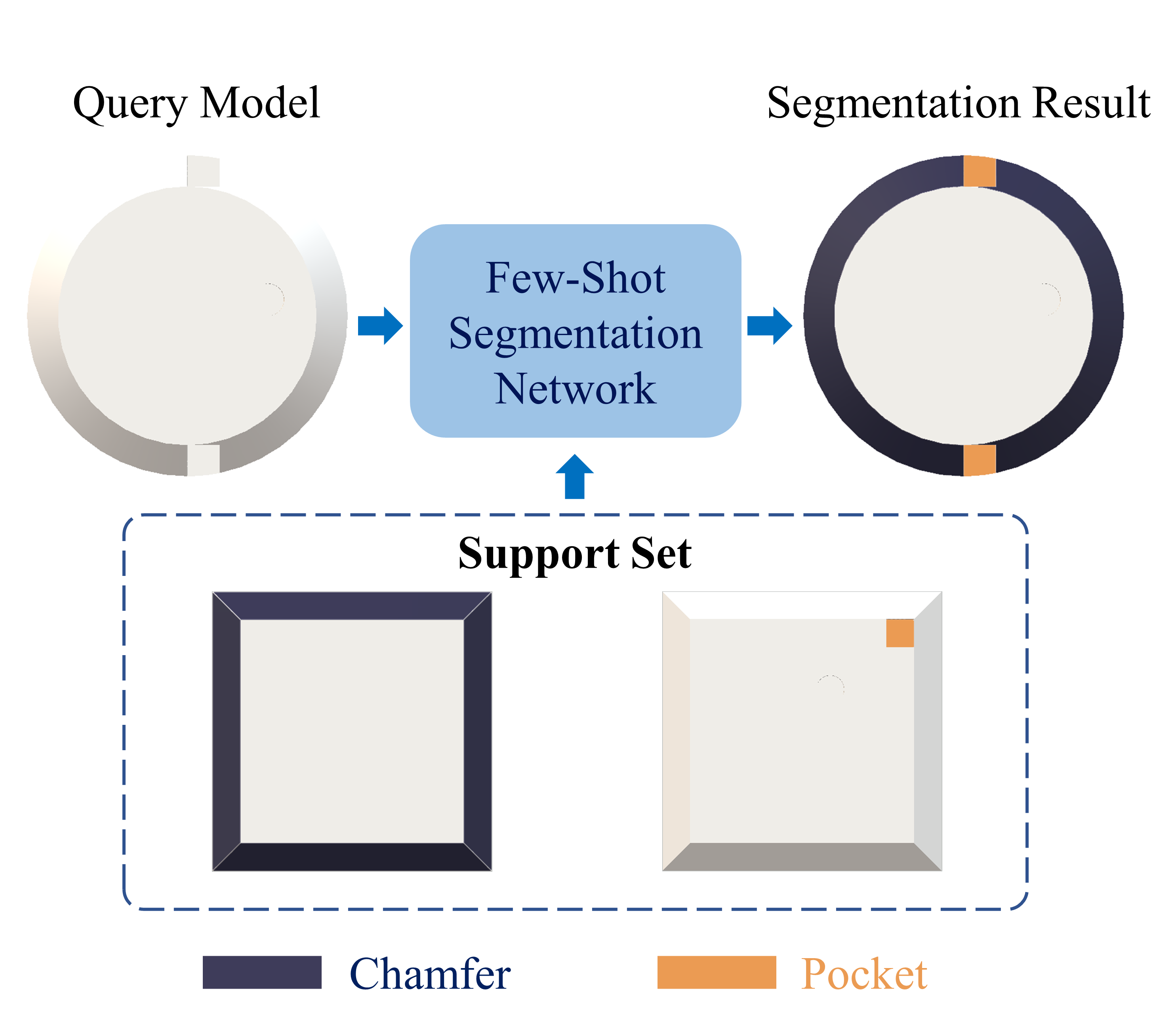}
\caption{The CAM/CAD models for support and query are transformed to point clouds, the trained few-shot segmentation network is able to recognize and segment the similar classes. The output point cloud can be reconstructed to CAM/CAD models in extra subsequent steps. This figure illustrates a segmentation result of 2-way 1-shot episode.}
\label{fig:fewshotillustrate}
\end{figure}

\begin{figure*}[htbp]
\centering
\includegraphics[width=18cm]{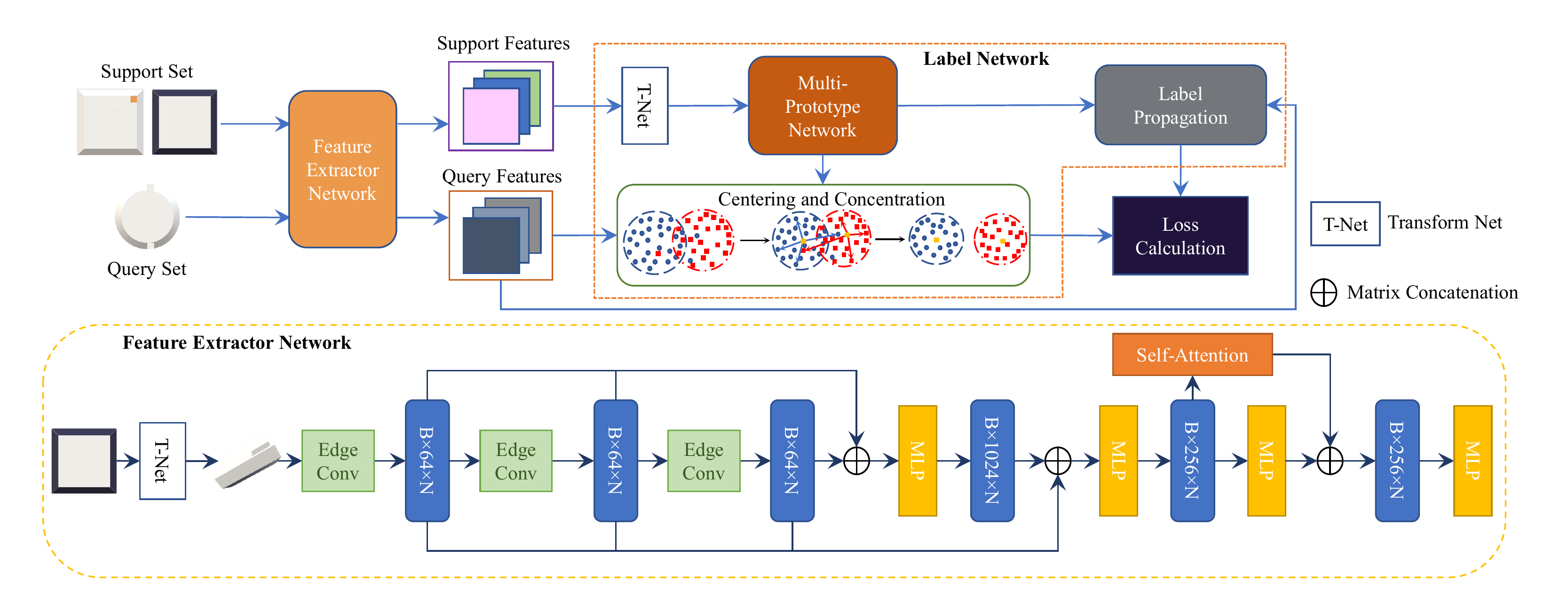}
\caption{The structure of our network which shows a 2-way 1-shot training. The Feature Extractor Network (FEN) maps input point clouds to feature space, the aim of label network is to label the query point cloud with label propagation algorithm, centering and concentration method is adopt for decrease the in-class distance of query features, loss function is calculated in the last block with the ground truth label of the query set as reference.}
\label{fig:netstructure}
\end{figure*}
\section{Introduction}
Computer Aided Manufacturing (CAM) is an essential step for modern manufacturing processes, 
where powerful professional-grade softwares are employed to conduct various series of steps to tell the machine 
how to manufacture the product automatically and effectively. 
Basically, the CAM workflow can be summarized as procedures involving checking geometric errors, 
generating tool path, setting tool parameters, and configuring offset of the machine\cite{CAMbook}. 
Among these work processes, recognition of 3D geometric features is a step of great importance. 
With the trend of miniaturization and integration, 
the manufacture of work-pieces requires more accurate 3D feature recognition. 
However, since there are all kinds of work-pieces with innumerable shapes, 
recognition of these features can turn out to be relatively difficult, and possibly exhausting too,
when attempted without appropriate and effective automation. 
Typically, traditional methods might not be able to recognize the features with satisfactory performance, 
while efforts in manually annotating the features are likely too difficult and inefficient. 
Without the automated capability of recognizing features efficiently and accurately, 
the manufacturing process will be slowed down;
and this
will also cause wastage of energy and time, 
which will further degrade the economic aspects of the associated manufacturing process.

CAM input usually has the same format as CAD models, 
and most traditional segmentation of these models are based on a so-called mesh method~\cite{CAMsegmethodTra,meshseg,meshseg2}, 
where the mainstream ideas are based on the notions of the mesh, 
and focuses on region-detection, 
clustering and multi-view projection~\cite{Sketch,cluster,region}. 
The major limitation in earlier more traditional methods (for recognition of 3D geometric features)
is that the 3D characteristic of the work-pieces cannot be fully enumerated 
and considered in the automatic algorithms. 
With the recent advances in computer vision~\cite{cv1,cv2} and applications of deep learning~\cite{dlapplication}, 
the possibility for new developments now arises where
the task of 3D feature detection 
can now be innovatively developed and
formulated 
as a vision task involving the notion of semantic segmentation. 
Recently too, it is well-known that deep learning 
can be successfully adopted 
to address the semantic segmentation problem in both 2D and 3D\cite{segdl,imgseg}. 

However, fully-supervised deep learning approaches require a complete large-scale dataset 
which is hard to achieve in 3D CAM/CAD. 
Moreover, most deep learning approaches address the segmentation tasks 
on a closed dataset where the part segmentation task must be trained in  fully-supervised manner. 
Although this can certainly improve the accuracy as the model has thus memorized the feature of the test part, 
nevertheless here,
the model lacks the ability to segment 
further additional
novel classes which are not similar with the training data (under the fully-supervised learning). 
Considering the diversity of work-pieces in the typical modern manufacturing workplace, 
practically thus (under the fully-supervised learning approach), 
it would then be not likely possible to effectively attain the required resulting AI model 
with that needed perfect and complete dataset that contains all possible features 
for CAM/CAD (encompassing all the possible 3D geometric features in an advanced typical modern manufacturing workplace).

To address the issues mentioned above, 
in this work here,
we propose a notable segmentation approach 
using few-shot learning; 
and this development
aims to effectively and automatically attain model segmentation 
for recognition of 3D geometric features of manufacturing parts,
and which can additionally incorporate
further additional novel classes. 
Along this line, it is pertinent to mention that
few-shot learning~\cite{fewshotconcept,fewshotobjdetect} is a well-regarded concept in the computer vision area. 
For this, the main idea is that after the model is trained on a relatively big dataset, 
the model is then also able to quickly learn with only a small number of samples for further new classes. 
Few-shot learning can be treated as the application of meta-learning in the field of supervised learning. 
Meta-Learning (also known as learning to learn), decomposes the dataset into different meta tasks 
in the meta training phase. 
This is carried out
to learn the generalization ability of the model 
for situations
when the category changes~\cite{metalearningconcept}. 
The aim in our development here is to train a model that can segment new work-pieces with a few labeled samples, 
as illustrated in Fig. \ref{fig:fewshotillustrate}. 
When we have a new work-piece, 
we can construct a few-shot task which is so called, an episode. 
An episode includes a query set that includes the capability to contain all new work-pieces, 
and a support set which includes trained and labeled data. 
The number of different classes in the support set is called ways of the few-shot task, 
and the number of samples of each classes are the shots. 
Fig. \ref{fig:fewshotillustrate} illustrates a 2-way 1-shot episode.

Generally, the few-shot learning methodology can be roughly divided into three categories, 
\textit{i.e.} 
metric-based methods~\cite{fewshotmetric1,fewshotmetric2}; 
model-based methods~\cite{fewshotmodel1,meta2}; 
and optimization-based methods~\cite{fewshotoptimal1,fewshotoptimal2}. 
Among these methods, metric-based methods are adopted in most efforts because of its simplicity and effectiveness. 
The goal of the metric-based method is to learn the similarity representation among part classes; 
as
parts that look like each other are supposed to have high similarity, 
while the parts that look different should have low similarity\cite{meta2}. 
The approach of the so-termed
prototype network\cite{prototype} is one of the rather more well-regarded metric-based methods. 
This maps each class to a prototype in feature space, 
and the metric of model is obtained by calculating the distance between test samples and prototypes.

In this paper, drawing inspiration from the pertinent and interesting work described as the attMPTI network\cite{attmpti}, 
we propose and develop a multi-prototype approach (with self-attention mechanics) for few-shot point cloud part segmentation. 
As mentioned, our work
aims to effectively and automatically attain model segmentation 
for recognition of 3D geometric features of manufacturing parts,
and which can additionally incorporate
further additional novel classes. 
Along this line, it is pertinent to note that
the developments described in attMPTI\cite{attmpti} focuses on indoor 3D point cloud semantic segmentation;
and this
pays more attention to the distribution of the points. 
However, 
part segmentation for work-pieces (for 3D CAM/CAD) 
also requires understanding of the 3D geometric features. 
Consequently, as an improvement and innovation, we additionally adopt a transform net in the embedding network 
to improve the comprehension for 3D features of the various possible instances of the whole work-piece. 
In addition, without attempting to invoke the need for the big datasets (with the attendant very large computational burdens etc.)
like s3dis\cite{s3dis} and scannet\cite{scannet}, shapenet\cite{shapenet}; 
the approach of a
part annotation dataset is used for pretraining, 
and a self-made small and relevant dataset 
(based on a plausible scenario of an essentially expected set of 3D geometric features of manufacturing parts) 
is used for training and testing.

\section{Our Method}

\subsection{Problem Formulation}
Recognition of 3D feature is an imperative step for CAM as many subsequent steps heavily rely on the feature recognition, \textit{e.g.}, tool path generation, parameters setting, and tool controlling~\cite{toolpathgen}. Considering the difficulty of label CAM models, building a data efficient deep learning network is beneficial and important. Therefore, 3D feature recognition is formulated as a few-shot learning problem to enhance the ability of segmenting novel classes without retraining and large dataset. As mentioned in previous section, we use episodic paradigm as the train and test format~\cite{episode}. The CAM/CAD models are transformed to point clouds as our dataset, the whole dataset $\mathcal{D}$ was split into training set $\mathcal{D}_{train}$ and test set $\mathcal{D}_{test}$. Assume that our dataset contain $N_{total}$ different classes, we randomly select $N_{train}$ classes for training and the remains $N_{test}=N_{total}-N_{train}$ classes are reserved for testing. Considering a $C-$way $K-$shot episode, we randomly select $C$ classes in $\mathcal{D}_{train}$, each class contain $K$ samples. Hence, the support set can be denoted as $\mathcal{S}=\{(\mathbf{D}_i^{1})_{i=1}^{K}\cdots (\mathbf{D}_i^{c})_{i=1}^{K}\}$ where $\mathbf{D}_i^{c}$ is the data pair that represents the $i$-th sample for $c$-th class, which contains a point cloud $\mathbf{P}^{c}_{i}$ and its point-wise label $\mathbf{L}^{c}_{i} \in \mathbb{R}^{n\times1}$ noticed that $\mathbf{P}^{c}_{i} \in \mathbb{R}^{n\times f}$, $n$ is the number of points in the point cloud and $f$ is the features of each point. In 
our datasets, $f = (x,y,z,X,Y,Z)$ which represents three dimension coordinates and three dimension normal vector respectively. Therefore, the support set can be written as $\mathcal{S}=\{(\mathbf{P}_i^{1},\mathbf{L}_i^{1})_{i=1}^{K}\cdots (\mathbf{P}^{c}_{i},\mathbf{L}^{c}_{i})_{i=1}^{K}\}$.  The query set $\mathcal{Q}= \{(\mathbf{P}^{i},\mathbf{L}^{i})\}_{i=1}^{N_{Q}}$ contain $N_{Q}$ samples that are randomly selected from $\mathcal{D}_{test}$. We additionally import a background class for each episode, consequently, the predict labels $\mathbf{Z} \in \mathbb{R}^{n\times(C+1)}$. Therefore, we can conclude our goal as train a model $M_{\Phi}(\mathbf{P}\in \mathcal{Q},\mathcal{S})$ with the optimal parameters $\Phi^{*}$ satisfied:
\begin{equation}
    \Phi^{*}=\underset{\Phi}{\operatorname{argmin}}\ \mathbb{E} \sim \mathcal{E}_{\text {train}}\left[\sum_{\mathbf{P}^{i}\in \mathcal{Q}} \mathcal{L}\left(\mathbf{L}^{i}, M_{\Phi}\left(\mathbf{P}^{i}, \mathcal{S}\right)\right)\right],
\end{equation}
where $\mathcal{E}_{\text {train}}$ is episode set sampled from $\mathcal{D}_{train}$, $\mathcal{L}$ is the loss function which will be detailed in the following sections.

\subsection{Transform Network}
In CAM process, work-pieces are mainly the combination of various geometries. Unlike scene scanning point clouds~\cite{s3dis,scannet} which have complex distributions, CAM/CAD segmentation is supposed to focus more on the vertex and edge of 3D work-pieces, \textit{i.e.} its topology. Consequently, to improve the model ability to comprehend the whole geometric feature of work-pieces, trainable transform net~\cite{pointnet} is used to grab the characteristic of point cloud that invariant to geometry transformation. 
The ideal orthogonal affine matrix $\mathbf{A}$ is able align point clouds in a specific view by matrix multiplying. The new point cloud $\mathbf{P^{\prime}}$ after transformation can be written as:
\begin{equation}
\begin{aligned}
&\mathbf{P^{\prime}}=\mathbf{\hat{A}}\mathbf{P}\\
&\mathbf{\hat{A}}=T_{net}(\mathbf{P}),
\end{aligned}
\end{equation}
where $\mathbf{\hat{A}}$ is the projection matrix predicted by $T_{net}$. In feature space, the affine matrix $\mathbf{\hat{A}}$ is to ensure the orthogonality a following regularization is imported:
\begin{equation}
    \mathcal{L}_{reg}=\left\|\mathbf{I}-\mathbf{\hat{A}} \mathbf{\hat{A}}^{T}\right\|_{F}^{2},
\end{equation}
where $||\cdot||_F$ refers to the Frobenius norm.
\subsection{Feature Extractor Network}
We employ the Dynamic Graph CNN (DGCNN)~\cite{dgcnn} as our backbone due to its ability to well understand the edge and vertex feature of a point cloud. The feature of the $i$-th point $\bm{p}_i$ in the point cloud with max pool operation can be defined as
\begin{equation}
    {V}_{i}=\varphi \left\{\max\limits_{\bm{p}_j \in \mathcal{K}(\bm{p}_i)} \big[\theta_{m}(\bm{p}_j-\bm{p}_i)+\phi_{m}(\bm{p}_i)\big]\right\},
\end{equation}
where $\mathcal{K}(\bm{p}_i)$ denotes the $k$ nearest neighbours of the point and $\bm{p}_i$ is the attributes of the point. $\phi_{m}$ and $\theta_{m}$ are parameters of the network, $\varphi$ is the activation function. Furthermore, we adopt self-attention unit and additional features due to the work-pieces usually contain small holes or pocket that global feature networks may not pay attention to. 

\subsection{Classification Head}
To obtain the final prediction of segmentation result, the query set point clouds should classified in point level and compare the ground truth label. Although parametric networks such as Multi-Layer Perceptron (MLP) can be chosen as the classifier, it will be seriously influenced by overfitting due to the small amount of samples in the query set and support set. In contrast, many non-parametric methods such as K-Nearest Neighbours(KNN), K-means and clustering do not need optimal parameters, and thus under meta-learning frameworks a few-shot classifiers can be trained end-to-end. This method is to model the distance distribution between samples, so that similar samples are close, and different samples are far away. Firstly we generate prototypes which can be treated as anchor points in feature space for each class in $\mathcal{S}$. Secondly, we adopt Label Propagation Algorithm (LPA) to label point clouds in the query set.

The multi-prototype network is developed based on prototype network~\cite{prototype}, it extends one prototype for each class to $n_{p}$ prototypes for each class. In our scenario, for each of $N$ class and a background class, we produce $n_{p}$ prototypes in feature space, we follow \cite{attmpti} which use the Farthest Point Sampling (FPS) algorithm to select $n_{p}$ anchor points and then arrange prototypes to each point based on the closest distance to prototypes. For instance, the set of prototypes for the $c$-th class  can be denoted as:
\begin{equation}
    \begin{array}{ll}
\boldsymbol{\mu}^{c}= & \left\{\boldsymbol{\mu}_{1}^{c}, \ldots, \boldsymbol{\mu}_{n_{p}}^{c} \mid \boldsymbol{\mu}_{i}^{c}=\frac{1}{\left|\mathcal{I}_{i}^{c *}\right|} \sum_{V_{j}^{c} \in \mathcal{I}_{i}^{c *}} V_{j}^{c}\right\} \\\vspace{1.2ex}
\text { s.t. } & \mathcal{I}^{c *}=\underset{\mathcal{I}^{c}}{\operatorname{argmin}} \sum\limits_{i=1}^{n_{p}} \sum_{V_{j}^{c} \in \mathcal{I}_{i}^{c}}\left\|V_{j}^{c}-{O}_{i}^{c}\right\|_{2},
\end{array}
\end{equation}
where $V_{j}^{c}$ and ${O}_{i}^{c}$ represents feature of the $j$-th point and the $i$-th anchor point in the point cloud respectively. $\mathcal{I}_{i}^{c}=\big\{\mathcal{I}_{1}^{c}, \ldots, \mathcal{I}_{n_{p}}^{c}\big\}$ is the subset we arranged to each prototypes.

In order to decrease the probability of misclassification and improve the ability of dealing with novel classes, we consider center loss\cite{centerloss} to reduce the in-class distance. Our aim is to ensure that the same class is close in the feature space. The center loss is defined as:
\begin{equation}
    \mathcal{L}_{c}=\frac{1}{2} \sum_{i=1}^{n\times B}\left\|V_{i}-\bm{\mu}_{y_{i}}\right\|_{2}^{2},
\end{equation}
where $n\times B$ represents the number of points in the batch, $B$ is the batch size, $V_{i}$ is the feature of the $i$-th point, $\bm{\mu}_{y_{i}}$ is the center of class ${y_{i}}$ in feature space. First, feature centers are updated in mini-batches instead of the entire training dataset. In each round of iteration, the average of each class feature is calculated; secondly, in order to avoid large loss fluctuation in the training process caused by some wrong samples, a hyper-parameter  $\lambda$, is used to control the update of the center loss which described in Eq.~\ref{eq:loss}.

The label propagation algorithm is a graph-based semi-supervised learning method, its basic idea is to predict the label information of unlabeled nodes from the label information of labeled nodes, and use the relationship between samples to build a complete graph model. The general process of it can be summarized as Algorithm~\ref{alg:lpa}.

\begin{algorithm}
\caption{General Label Propagation Algorithm} 
\begin{algorithmic}[1]
\REQUIRE  $n_{l}$ labeled data, $n_{ul}$ unlabeled data, epoch number $m$
\STATE Calculate weights between $V_i$ and $V_j$\\\vspace{0.5ex} $w_{i j}=\exp \left(-\frac{d_{i j}}{\sigma^{2}}\right)=\exp (-\frac{\sum_{d=1}^{N}\left(V_{i}^{d}-V_{j}^{d}\right)}{\sigma^{2}})$\vspace{1ex}
\STATE Define propagation matrix $\mathbf{T} \in \mathbb{R}^{(n_{l}+n_{ul})*(n_{l}+n_{ul})}$\vspace{1ex}
\STATE Calculate probability of label $j$ propagates to label $i$\\\vspace{0.5ex} $\mathbf{T}_{i j}=P(j \rightarrow i)=\frac{w_{i j}}{\sum_{k=1}^{n_{l}+n_{ul}} w_{k j}}$\vspace{1ex}
\STATE Define class matrix $\mathbf{Y} \in \mathbb{R}^{(n_{l}+n_{ul})*(N+1)}$ where $N$ is the all classes number. Rows of $\mathbf{Y}$ represent labeled data are one-hot labels and others are zero initially.
\WHILE{epoch$<$$m$} 
\STATE Propagate label with $\mathbf{Y}_{i,n}=\sum\limits_{j}\mathbf{T}_{ij}*\mathbf{Y}_{j,n}\ j \in \mathcal{K}(i)$
\ENDWHILE
\ENSURE $n_{l}+n_{ul}$ labeled data
\end{algorithmic}
\label{alg:lpa}
\end{algorithm}

The label of each node is propagated to adjacent nodes according to the similarity. At each step of node propagation, each node updates its label according to the label of the adjacent node. The greater the similarity with the node, the larger the influence weight of its neighboring nodes on its labeling. The more consistent the labels of similar nodes are, and the easier its labels are to propagate. During label propagation, the label of the labeled data is kept unchanged, so that it passes the label to the unlabeled data. Finally, when the iteration ends, the probability distributions of similar nodes tend to be similar and can be classified as one class. In our paper, we follow\cite{lp1} to update the label:
\begin{equation}
    \mathbf{Z}_{t+1}=\alpha {\mathbf{D}^{-1/2}(\mathbf{W}+\mathbf{W}^{T})\mathbf{D}^{-1/2}} \mathbf{Z}_{t}+(1-\alpha) \mathbf{Y},
\end{equation}
where $\mathbf{W}$ is the weights matrix, $\mathbf{D}$ is the diagonal matrix whose value are sum of rows of $\mathbf{W}$. $\alpha$ is probability parameter which usually are set between 0 and 1. We follow\cite{lp2} to get the optimal label matrix.
\begin{equation}
    \mathbf{Z}^{*}=(\mathbf{I}-\alpha (\mathbf{D}^{-1/2}(\mathbf{W}+\mathbf{W}^{T})\mathbf{D}^{-1/2}))^{-1} \mathbf{Y}.
\end{equation}

The prediction of probability for the $i$-th point belong to class $c$ is denoted as $\bm {Z}_{i}^{c}$, the final predicted class probability is expressed as
\begin{equation}
    \bm{H}_{i}^{c}=\frac{\exp \left(\bm{Z}_{i}^{c}\right)}{\sum_{j=1}^{N+1} \exp \left(\bm{Z}_{i}^{j}\right)}.
\end{equation}
The loss function of this block is based on cross-entropy loss, by minimizing the loss function we are actually maximizing the lower boundary of the mutual information between prediction and ground truth.
\begin{equation}
    \mathcal{L}_{m}=-\frac{1}{|\mathcal{Q}|} \frac{1}{n} \sum_{i=1}^{|\mathcal{Q}|} \sum_{j=1}^{n} \sum_{c=1}^{N+1} \delta\left(\bm{L}_{j}^{i},c\right) \log \left(\bm{H}_{i, j}^{c}\right),
\end{equation}
where the ground truth label $\bm{L}_{j}^{i}$ which represents the $j$-th point in the $i$-th point cloud of query set and function $\delta(\cdot)$ is defined as:
\begin{equation}
    \delta(x,y)=\left\{
\begin{aligned}
1\quad & x=y \\
0\quad & x\ne y \\
\end{aligned}
\right.
\end{equation} 

\begin{figure}[tbp]
    \centering
    \includegraphics[scale=0.36]{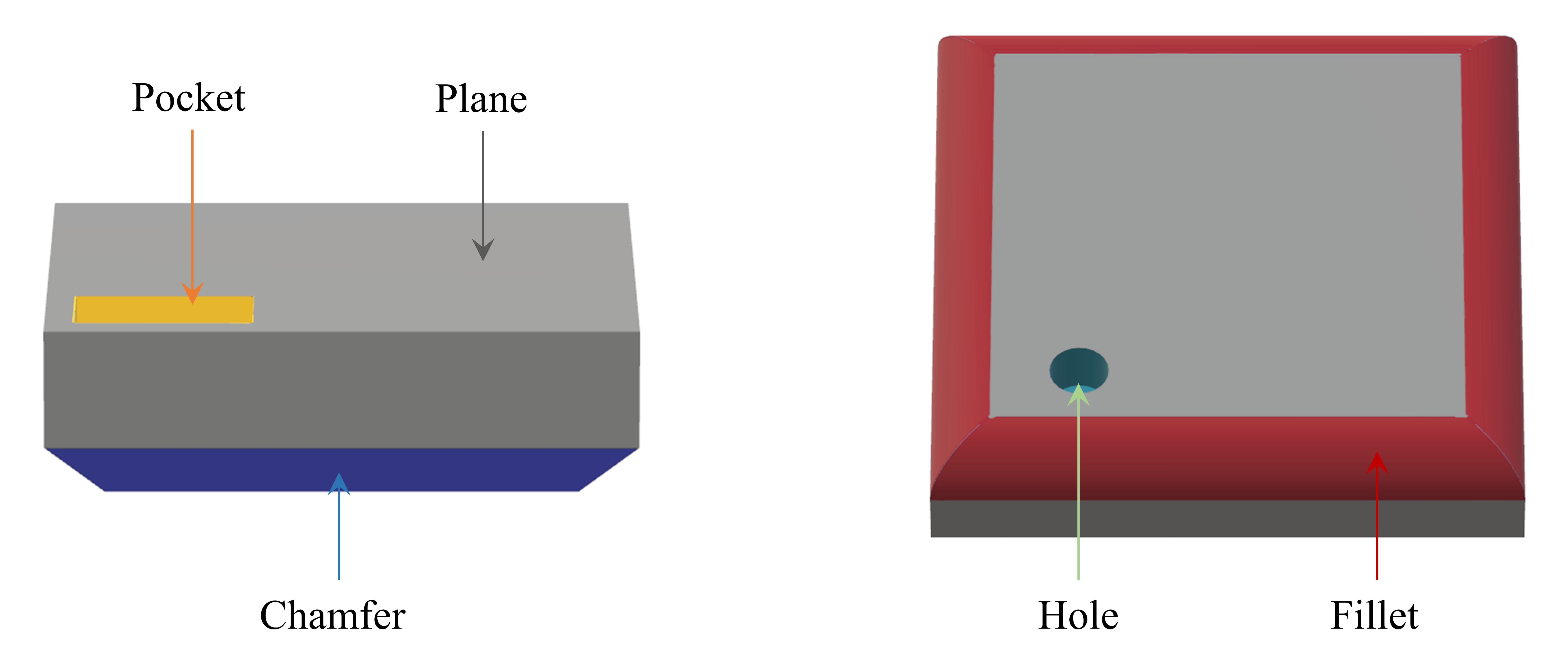}
    \caption{The example of two work-pieces that include five classes in the dataset}
    \label{fig:classexample}
\end{figure}
\subsection{Learning Strategy}
Pretraining and fine-tuning strategy is simple yet effective~\cite{pretrain,pretrain2}, we follow this mode and our loss is the sum of the previous three losses, \textit{i.e.},
\begin{equation}
  \mathcal{L}=\mathcal{L}_{m}+\lambda\mathcal{L}_{c}+\mathcal{L}_{reg},
  \label{eq:loss}
\end{equation}
In pretraining, the FEN is trained on Shapenet dataset~\cite{shapenet}. During this stage, the sum of cross-entropy loss $\mathcal{L}_{m}$ and regularization $\mathcal{L}_{reg}$ is adopt as the pretrain loss. By pretraining the FEN, we will have better initialization for feature space distribution in training. As for training, the fine-tuning baseline connects a MLP as the classification head and fine-tunes the FEN and the classification head. We use the label network which is shown in Fig.~\ref{fig:netstructure} to label and concentrate the samples in query set $\mathcal{Q}$ and then use MLP with softmax function to make final predictions. In the training stage, the gradient of the loss function will also flow into the FEN, to avoid significant changes in our FEN, we fixed all parameters in the FEN except the last two MLP layers. Noticed that, the cross-entropy loss was calculated directly by the ground truth label of point clouds and the prediction from our network.

\section{Experiments}
\subsection{Datasets Preparation}
We split our processes into pertraining and training. In pretraining we use Shapenet\cite{shapenet} part annotation subset as our dataset which contains 16,849 point clouds. For training stage, we use dataset in \cite{caddataset} which is obj files of CAD models. We sampled 5,000 points for each CAD model to construct datasets using an open source software cloudcompare which adopt Poisson disk as sampling algorithm\cite{cloudcompare}. We manually select and labeled 2,400 typical point clouds correspond to five classes as our datasets. To identify the training set and test set, we use cross-validation strategy among two subsets which are split on account of different classes.

We follow \cite{dgcnn} and randomly select 2,048 points for each point cloud. To generate an episode in training, we randomly choose one subsets as $\mathcal{D}_{train}$ then select $C_{train}$ classes for support set $\mathcal{S}$ and query set $\mathcal{Q}$. For an episode in testing, we randomly select classes in the rest, support set and query set are constructed by iterate combination of samples. Fig. \ref{fig:classexample} illustrates the five classes in our datasets \textit{i.e.} Hole, Pocket, Chamfer, Fillet and Plane. In practical segmentation, the Plane class is treated as background when the segment goal is a combination of other four classes.

\subsection{Implementation Details}
We follow\cite{attmpti} to set the parameters of the label propagation and multi-prototype blocks. $\lambda$ is set to 0.9 to guarantee the cluster performance. In the pretraining stage, the feature extractor are trained on shapenet dataset with 32 point clouds per batch and 200 epoch with total 105,400 iterations. Because of our small training dataset, the batch size is set to 16 during training stage, the total number of training iterations are set to 25,000. The first baseline is adopt from fine-tuning networks (FTN)\cite{finetune}. For comparison, DGCNN vanilla is used for segmentation, it is trained on point clouds in all support sets and tested on all query sets. Therefore, it will show equal result in different few-shot settings. In training stage, we generate 100 episodes to train the network. Noticed that a single point cloud can be shown in different support set as representative of different classes, because point clouds in support set only have the label of represented class. Five classes are split to 2 groups for training and testing respectively, if one of them was used for training, the other will be used for testing. 
\begin{figure}[tbp]
    \centering
    \includegraphics[scale=0.301]{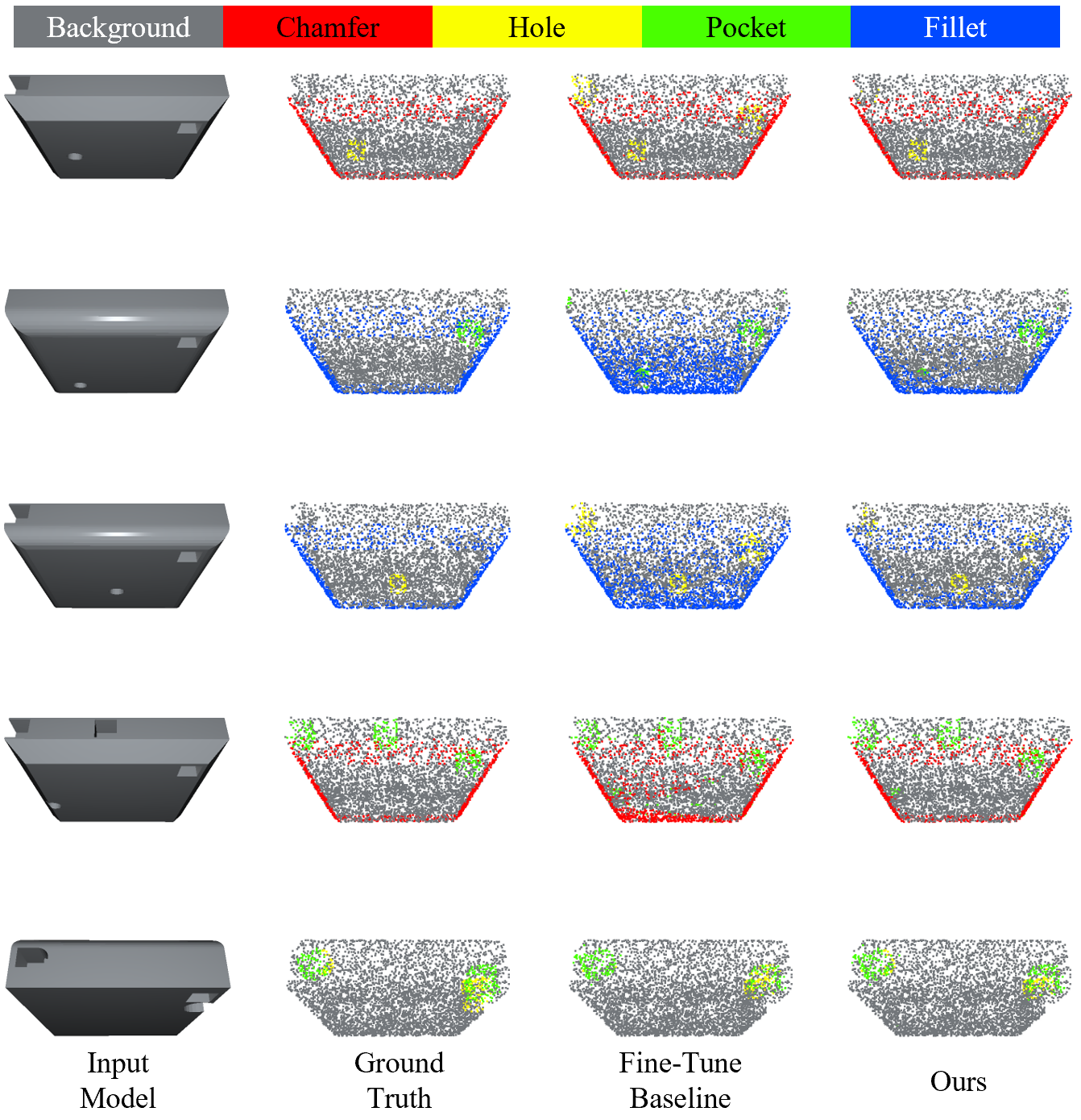}
    \caption{Segmentation result of 2-way 1-shot model on \textbf{square work-pieces}. Each row represents a combination of 2-way episodes \textit{i.e.}, ``\textit{Chamfer,Hole}" for the first row, ``\textit{Fillet,Pocket}" for the second row, ``\textit{Fillet,Hole}" for the third row, ``\textit{Chamfer,Pocket}" for the fourth row, ``\textit{Hole,Pocket}" for the last row.}
    \label{fig:res}
\end{figure}
\subsection{Result and Analysis}
\begin{table}[htbp]
    \caption{Results of different methods using mIoU as criteria (\%).}
    \begin{center}
    \begin{tabular}{|c|c|c|c|c|c|c|}
    \hline
    & \multicolumn{2}{|c|}{\textbf{2-way 1-shot}}& \multicolumn{2}{|c|}{\textbf{2-way 3-shot}}& \multicolumn{2}{|c|}{\textbf{2-way 5-shot}} \\
    \cline{2-7} 
    \textbf{Method} & Best & Mean& Best & Mean&Best & Mean \\
    \hline
    DGCNN & 21.20 &  19.58 & 21.20 & 19.58 & 21.20 & 19.58 \\
    \hline\vspace{0.2ex}
    FTN & 41.54 &  39.01 & 51.29 & 50.32 & 54.41 & 52.93 \\
    \hline
    \makecell{Without \\Center Loss}  &46.92 &44.35 &52.17 &51.40 &55.21 &53.80 \\
    \hline
    Without T-Net & 48.19 &45.51 &55.35 &53.94 &57.60 &56.29\\
    \hline
    \textbf{Ours} & \textbf{48.28} & \textbf{46.88} & \textbf{55.90} & \textbf{54.82} & \textbf{57.74} & \textbf{57.37} \\
    \hline
    \end{tabular}
    \label{tb:expresult}
    \end{center}
\end{table}
After test 5 times on each  methods, Table \ref{tb:expresult} shows the best and mean mIoU respectively. The DGCNN vanilla, no surprise,  performed relatively bad on the task because of overfitting. The fine-tuning baseline and our model have higher mIoU. There is no doubt that, with the larger shot number $K$, the mIoU will be higher, because the model has more references in the support set. As shown in the table, the improvement from one-shot to three-shot is larger than what from three-shot to five-shot. This might because the dataset we choose is relatively simple, the model can learn the geometric features of the data well, but overfitting becomes a problem in this case.Without the T-Net, the best mIoU is relative similar to our method while the mean mIoU are lower than ours this is because of there is slight overfitting when we abrogate the T-Net. Without the center loss, the model performs worse due to higher misclassification probability which demonstrate the advantages of our method relatively.

\begin{figure}[tbp]
    \centering
    \includegraphics[scale=0.31]{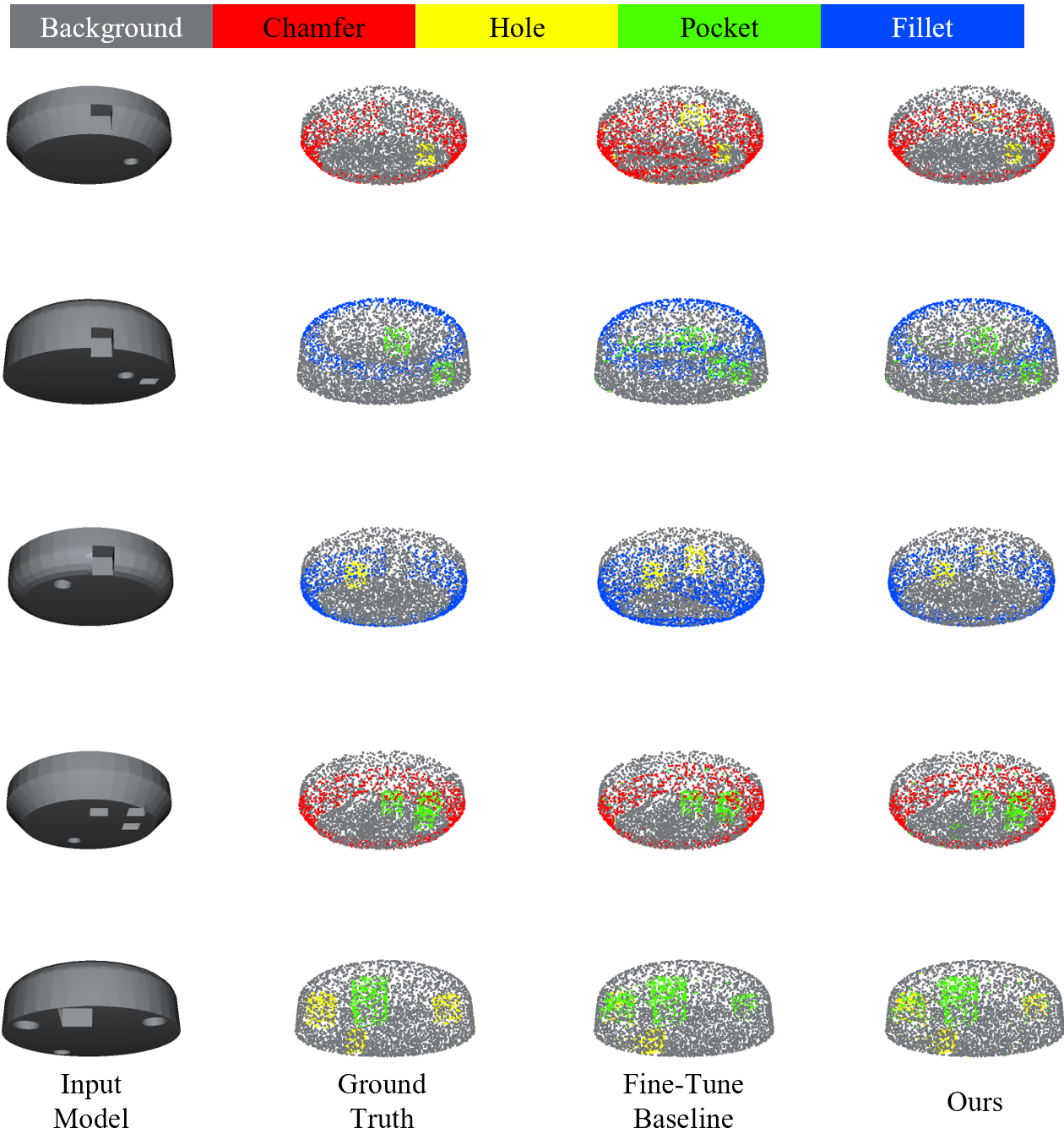}
    \caption{Segmentation result of 2-way 1-shot model on \textbf{round work-pieces}. Each row represents a combination of 2-way episodes \textit{i.e.}, ``\textit{Chamfer,Hole}" for the first row, ``\textit{Fillet,Pocket}" for the second row, ``\textit{Fillet,Hole}" for the third row, ``\textit{Chamfer,Pocket}" for the fourth row, ``\textit{Hole,Pocket}" for the last row.}
    \label{fig:res2}
\end{figure}

In addition, to improve the performance, the number of shots should be considered carefully, we can also choose new regularization functions or different learning processes such as contrastive learning which will enhance the generalization ability of the model. Fig.\ref{fig:res} and Fig.\ref{fig:res2} shows the segmentation result of square work-pieces and round work-pieces respectively.

For fillets, the segmentation result is worse than chamfers. One possible reason is, comparing to chamfers, fillets do not have sharp geometric boundaries with the plane. This also can explain why the segmentation result of pockets is better than holes. In other words, models are more likely to recognize the sharp geometric features. We assume that with better generalization ability and learning process with data augmentations, the model may able to recognize the smooth geometric features as well.
\section{Conclusion}
In this work, we propose and develop a noteworthy few-shot learning-based approach 
for effective part segmentation in CAM/CAD. 
This is designed 
to significantly enhance its generalization ability,
and our development also
aims to flexibly adapt to new segmentation tasks by using only relatively rather few samples. 
A novel few-shot solution is proposed based on the center concentrated multi-prototype transductive inference method. 
Moreover, the methodology of a transform net is adopted to improve the geometry invariance of feature extraction 
in
understanding the 3D work pieces. 
Automated part segmentation based on our method 
can achieve the greatly improved levels of consistency and accuracy 
for which traditional methods will encounter difficulty to reach. 
Finally,
our approach stores data in the point cloud format 
that more straightforward for understanding, 
and which also makes 
the various procedures involved 
have significantly
easier read and edit access (thus improving efficiency and effectiveness and lowering costs).


\section*{Acknowledgment}
This research is supported by the National University of Singapore under 
the NUS College of Design and Engineering Industry-focused Ring-Fenced PhD Scholarship programme. 
It is also supported in part by Singapore Institute of Manufacturing Technology (SIMTech), A*STAR.
Additionally,
the authors would like to acknowledge useful discussions with Dr.Hendrik Schafstall of
Hexagon, Manufacturing Intelligence Division,
Simufact Engineering GmbH.

\end{document}